\documentclass[aps,pre,twocolumn,superscriptaddress]{revtex4-2}

\usepackage{amsmath}
\usepackage{amssymb}
\usepackage{comment}
\usepackage{amsthm}
\usepackage{tikz}
\usepackage{graphicx}
\graphicspath{ {Images/} }
\usepackage[T1]{fontenc}
\usepackage[utf8]{inputenc}
\usepackage{hyperref}
\hypersetup{colorlinks=true,citecolor=blue,linkcolor=blue,urlcolor=blue}

\begin{document}

\title{Congestion and extreme events in urban street networks}
\author{Ajay Agarwal}
\email{ajayagarwal28031998@gmail.com}
\affiliation{Department of Physics, Indian Institute of Science Education Research, Pune 411008, India.}

\author{M. S. Santhanam}
\email{santh@iiserpune.ac.in}
\affiliation{Department of Physics, Indian Institute of Science Education Research, Pune 411008, India.}

\date{\today}

\begin{abstract}
Congestion and extreme events in transportation networks are emergent phenomena with significant socio-economic implications. In this work, we study congestion and extreme event properties on real urban street (planar) networks drawn from four cities and compare it with that on a regular square grid. For dynamics, we employ three variants of random walk with additional realistic transport features.  In all the four urban street networks and 2D square grid and with all dynamical models, phase transitions are observed from a free flow to congested phase as a function of birth rate of vehicles. These transitions can be modified by traffic-aware routing protocols, but congestion cannot be entirely mitigated. In organically evolved street networks, we observe a semi-congested regime which has both congested and free-flow components. In the free-flow regime, the extreme event occurrence probability is larger for small degree nodes than for hubs, a feature originally observed in non-planar scale-free networks. In general, with respect to congestion and extreme events, the urban street networks and regular square grid display similar properties.
\end{abstract}

\maketitle

\section{Introduction}
Extreme events are often associated with various disasters such as floods, forest fires, heatwaves, volcanic eruptions, earthquakes and power-grid failures \cite{albeverio2006extreme}. They occur rarely and are characterised by strong deviations from mean behaviour. When any dynamical variable, as a function of time breaches a tolerance limit, due to some inherent fluctuations, then it is considered to be an extreme event \cite{santhanam2008return,kishore2011extreme,kishore2012extreme,santhanam2012extreme,kishore2013manipulation,kumar2020extreme,GupSan2021, gandhi2022biased}. For instance, suppose $x(t)$ represents the number of vehicles passing through a road junction at time $t$, an extreme event is said to occur if $x(t) > q$, where $q$ is some threshold usually related to the servicing capacity of the road junction. 

While extreme value theory was well-studied for the last one century \cite{coles2013}, extreme events taking place on a topology of networks has attracted considerable attention only in the last one or two decades \cite{MasPorLam2017,NagRayDan2022}. The latter is motivated by real-life phenomena, {\it e.g.,} blackouts in power-grids, traffic jams, congestion on road networks and IP network congestion, and ultimately to questions about the resilience and robustness of networks against failures \cite{artgraDe2024,LiuLiMa2022,ValSheLa2020,SpiSolFor2018,ves2010,KorBarTuc2018,YuaShaSta2015}. Usually, these questions are addressed by applying localized perturbation on networks and studying its cascading features \cite{KorBarTuc2018,artgraDe2024}. One approach to studying extreme events on networks is by employing many independent random walkers, whose dynamics defines a flow on the network, and extreme events emerge due to purely intrinsic fluctuations \cite{kishore2011extreme,noh2004random}. In this scenario, an ``event'' refers to the instantaneous number of walkers at a node. Such an event is designated as {\it extreme event} if the number of walkers $w(t)$ at time $t$ exceeds a pre-defined threshold $x_{\rm th}$. In this setting, it is remarkable that small degree nodes are more prone to EEs (on an average) than the hubs which attract large flux but are less prone to EEs \cite{kishore2011extreme}. Subsequently, this general feature was shown to hold good for extreme events in different scenarios and systems: in shortest path and biased random walks on networks \cite{kishore2012extreme}, overload failures arising from fluctuating loads \cite{ShoKou2015}, the dynamics of Brownian particle in an external potential \cite{AmrMaHu2018}, and in target routing model of non-interacting packets on a network \cite{LinHuDin2013}. Depending on the bias applied to the random walkers, extreme events {\it on edges} of a network also displayed dependence on local properties of the edge, though different from the general result described above \cite{gandhi2022biased}.

However, one important exception was reported for the flow defined in terms of Nagel-Scheckenberg model for road traffic on a small {\it synthetic network} with 30 nodes \cite{GupSan2021}. In this case, the extreme event occurrence probability appeared to be uniformly distributed over degree of the nodes on the network. In general, not much attention has gone into focussing on extreme event related questions for an important application, namely, the traffic on road networks. Another phenomenon of significant interest is the network traffic congestion \cite{GanKitMar2017, hull2004mitigating, jacobson1988congestion, huberman1997social, wang2009abrupt, wang2006traffic, de2009minimal, de2009congestion} on IP and road networks that leads to slow mobility, longer trip times, and increased queue lengths, and ultimately to even dysfunctional network. In the present context, extreme events relate to breaching a threshold value in steady state dynamics, while congestion refers to a dynamical state characterized by the failure of network to support flux on some part or all of its nodes. A question of interest is whether there can exist a transition from free-flow state to congested state on networks, and if so, how does it depend on network structure and dynamics. Many earlier studies have dealt with this question using synthetic networks, while we obtain results on real urban street networks. Apart from network topology, inefficient routing protocols may give rise to congestion in the system. Hence, a realistic routing protocol that can display free flow and congested states, and transitions between them will be useful. 

To study these questions about congestion and extreme events in a realistic setting, in this work, we evaluate a realistic model of ``vehicle'' dynamics taking place on real urban road networks drawn from four big cities \cite{CcoLimGon2016}. Towards this end, we introduce and employ a variant of the traffic model first studied in Ref. \cite{de2009minimal, de2009congestion}. Unlike the standard random walk dynamics \cite{noh2004random}, this model allows for total number of vehicles on the network to vary with time, and integrates crucial elements derived from real-world schemes while striking a balance between topology-based and traffic-based routing strategies. The dynamics induced by this model features a free-flow and congested regime for traffic depending on the parameters being chosen, and also transitions between these regimes. Further, it can reproduce the observed crossover in traffic fluctuation scaling within the Internet \cite{echenique2004improved, echenique2005dynamics, barankai2012effect, pastor2004evolution, pastor2001epidemic, hull2004mitigating, de2004adaptive}. As for networks, we employ real road networks of large urban cities -- namely, Ahmedabad, Mumbai, New Delhi (all in India), and New York (in USA) -- drawn from a public dataset \cite{osm_web}. 

\begin{figure}[t]
    \centering
    {\includegraphics[width=0.35\textwidth]{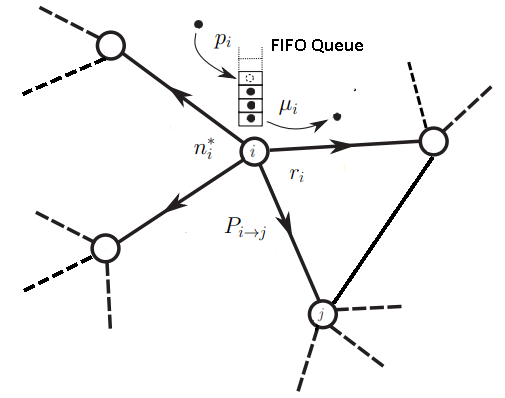}}
    \caption{A schematic of the transport model on network. For a node labelled $i$, equipped with a FIFO queue, the parameters that characterize the dynamics are as follows: $n_i^*$ (capacity of the FIFO queue), $p_{i}$ (particle creation rate), $r_{i}$ (service rate or outflux), $\mu_i$ (particle absorbing rate), $P_{i \rightarrow j}$ (transition probability from node $i$ to $j$).}
    \label{fig:Queue_network}
\end{figure}

For comparison and benchmarking purposes, we also use a synthetic 2D square grid lattice. Even though the microscopic details of road network of these 4 cities are different, we show that the extreme event properties are nearly the same on all of them and, most importantly, are well captured by that on a square grid lattice. Thus, at least as far as extreme events are concerned, realistic effects are not too different from that emerging on a square lattice. Further, from a statistical physics point of view, this work will also be of interest as an exploration of extreme event properties on disordered nearly-planar network.

In the next section, our traffic model and sources of urban street network data are described. In Sec III, the results for three variants of traffic model on street networks are described and compared with than on 2D regular square grid. Finally, in Sec IV, we provide a summary and discussion of the central results of the paper.

\begin{figure}[t]
    \includegraphics[width=0.35\textwidth]{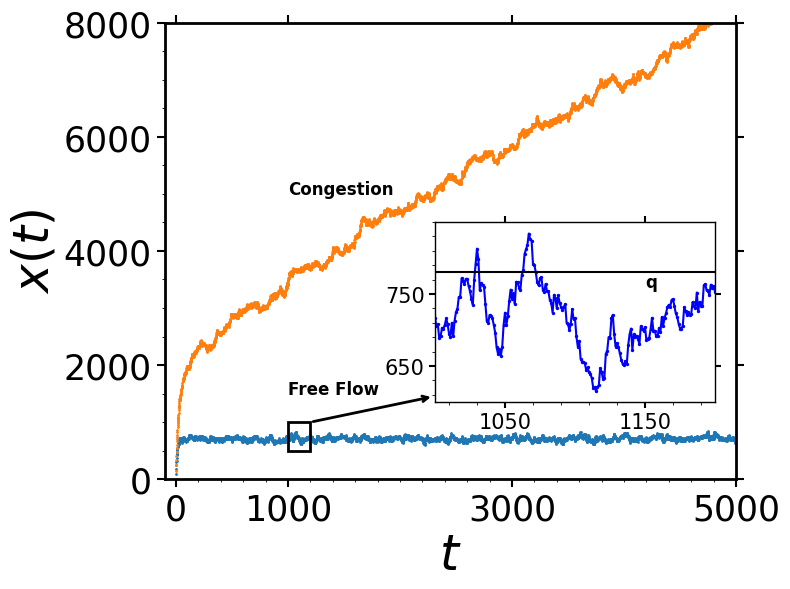}
    \caption{Congestion and free flow states. In congested state, the flux of vehicles $x(t)$ continuously increases with time, resulting in long queues. In free-flow state, queues are regularly cleared in time scales of $\Delta t \ll T$, where $T$ is the total simulation time, causing $x(t)$ to fluctuate around a mean value. Inset: illustrates extreme events. An event is designated as an instantaneous extreme event if $x(t)>q$, where $q$ is the threshold (black line).}
    \label{fig:TimeS}
\end{figure}

\section{Transport Model and Urban Street Network}
{\it Transport model}: We employ a variant of the traffic model studied in Ref. \cite{de2009minimal, de2009congestion}. A schematic of the transport model is shown in Fig. \ref{fig:Queue_network}. Consider a network with $N$ nodes and \{$\Omega_i$\} represents the set of neighbours of node $i$. At every time step, particles are created at each node with a probability of $p_i$. Thus, the dynamics of particles can be regarded as a continuous time stochastic process. Each node is endowed with a First-In First-Out (FIFO) queue in which arriving particles are stored. Let $n_i$ be the number of particles in the queue of node $i$. At each time step, in the node for which $n_i > 0 $, $r_i$ particles leave $i$-th node and jump into the queue of a randomly chosen neighbour $j \in \{\Omega_i\}$. The arriving particle can be rejected by the node $j$ with probability $\eta (n_j)$ and then the particle remains on the node $i$. If not rejected, the particle is either destroyed during the hopping with a probability $\mu_j$ or enters the queue of node $j$ with probability $1-\mu_j$. For simplicity, we consider 
\begin{equation}
    p_i=p, \;\;\;\;\;\; \mu_i = \mu, \;\;\;\;\;\; \eta(n_i) = \bar{\eta} ~ \theta(n_i-n_i^\ast),
\end{equation}
where $\theta(x)$ is the Heaviside step function. Here, $\eta(n_i)$ is a congestion control parameter, by varying $\bar{\eta}$ different routing protocols can be realized. In this work, we study this model and its variant (described in Sec III) on real street networks. 

To assess the efficacy of the protocol in ensuring that all packets generated at a rate of $Np$ successfully reach their destination within the shortest possible duration, it is imperative to evaluate whether the network can effectively handle the influx of all incoming vehicles or flux. If the network possess the capability to accommodate this influx, the total number of vehicles within the system at time $t$, denoted by $A(t) = \sum_i^{N}n_i$, will remain constant over time, indicative of a stationary state or free-flow regime. Conversely, if the network fails to manage the incoming traffic adequately, $A(t)$ will exhibit a continuous increase over time, signaling the onset of congestion within the network.
In order to distinguish between the free-flow and congested phases, we use the order parameter $\rho$ \cite{arenas2001communication, arenas2003search} defined as
\begin{equation}
\rho = \lim_{t \to \infty} \frac{A(t+\tau) - A(t)}{Np\tau}.
\label{eq:orderparam}
\end{equation}
In this, $Np$ is the average number of particles created at each time step and $\tau$ is the observation duration. Note that $p = \frac{\sum_i p_i}{N}$ is the average birth probability. The system attains a congested state if the order parameter exceeds $\rho > 0$. Larger values of $\rho$ correspond to severe congestion with $\rho = 1$ indicative of a highly congested state, while $\rho = 0$ signifies an absence of congestion implying a free-flow scenario.

\begin{table}[t]
    \centering
    \begin{tabular}{|c|c|c|c|c|c|}
         \hline
         City  & Radius & $N$ & $E$ & Total road\\
         (Reference coordinate)  & (Kms) &              &              & length (Kms)\\
         \hline
         \hline
         Ahmedabad &  1.5 & 1064 & 1528 & 135.3 \\
         (23.03$^{\circ}$N, 72.58$^{\circ}$E) & & & & \\
         \hline
          NewYork &  2 & 1117 & 1930 & 193.3 \\
          (40.71$^{\circ}$N, 74.01$^{\circ}$W) & & & & \\
         \hline
         Delhi &  0.95 & 1092 & 1708 & 99.5 \\
         (28.70$^{\circ}$N, 77.10$^{\circ}$E) & & & & \\
         \hline
         Mumbai & 2.15  & 1071& 1522 & 169.7 \\
         (19.07$^{\circ}$N, 72.87$^{\circ}$E) & & & & \\
         \hline 
    \end{tabular}
    \caption{Basic information about street networks of Ahmedabad, Delhi and Mumbai (all in India), and New York (in USA). Radius refers to a notional cirlce centred at the reference coordinate. The street network within this radius is considered for analysis in this work. Here, $N$ and $E$ are the number of nodes and edges, respectively, enclosed within the area considered for analysis.}
    \label{tab:tab1}
\end{table}

Figure \ref{fig:TimeS} displays results of simulating this transport model on a network of 1092 nodes. For values of parameters $\mu_i = 0.2$, $\bar{\eta}=0$, $n_i^{\ast} = 10$ and $r_i = 1$ for all $i=1,2, \dots N$, the temporal dynamics of both the congested state and free-flow state corresponding to $p=0.15$ and $p=0.10$ respectively, are depicted. In the congested state $A(t)$ grows in an unbounded manner, while for the free-flow state $A(t)$ does not grow on an average. In a congested state, due to unbounded increase in $A(t)$, network will cease to support flux. On the other hand, a physically interesting situation corresponds to extreme events, and this is studied in the free-flow regime by applying a threshold as depicted in the inset of Fig. \ref{fig:TimeS}. In this regime, there can be occasional spurts of extreme events but they may not always affect the entire network as a whole.

{\it Street Network data}: Spatial networks of urban streets refer to the interconnected system of roads, streets, and pathways that comprise the transportation infrastructure of any urban settlement. These networks can vary in complexity, with some cities having a more grid-like pattern while others have a more organic and meandering layout.
For instance, street network in parts of New York is known for its grid-like pattern while Ahmedabad, Delhi and Mumbai, on the other hand, have more organic and unplanned street networks that have developed over time. 

\begin{figure*}[t]
    (a)\includegraphics[width=3.3cm]{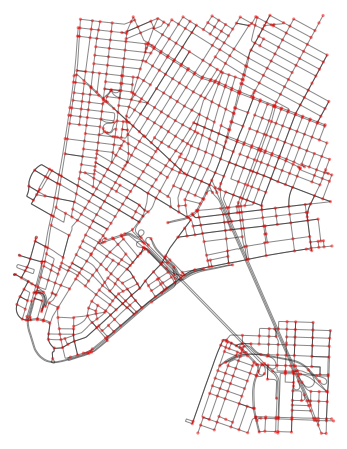} \hfill
    (b)\includegraphics[width=4.2cm]{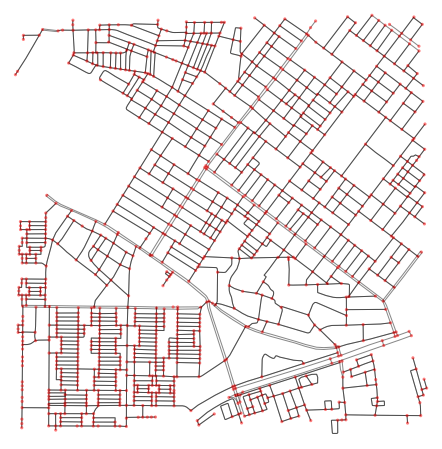}  \hfill
    (c)\includegraphics[width=4.2cm]{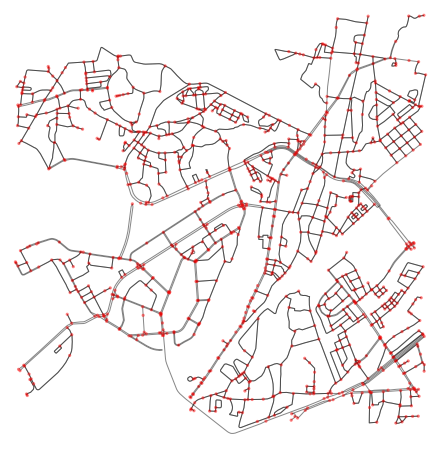}   \hfill
    (d)\includegraphics[width=4.2cm]{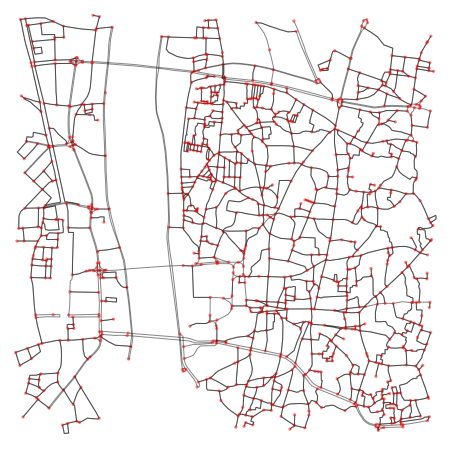}
    \caption{Spatial network of urban streets of selected region (see Table \ref{tab:tab1}) in (a) New York, (b) Delhi, (c) Mumbai and (d) Ahmedabad. The red dots represent nodes and grey lines the edges. }
    \label{fig:spat}
\end{figure*}

In this work, we use the urban street networks of major Indian cities, including Delhi, Mumbai, and Ahmedabad (that had organic evolution over time) and compare it with planned street network of New York, USA. The spatial network data, denoted by $G(N, E)$, is obtained from {\tt openstreetmap.org} \cite{osm_web} using python-based package OSMnx \cite{software1}. Here, $N$ and $E$ represent the number of nodes and edges in the network, respectively. 
To capture approximately 1000 nodes for our analysis, we select a region with a specific radius around a reference coordinate. The data structure of networks obtained using OSMnx is a multi-digraph. However, for ease of analysis, we converted it into a simple undirected network. Table \ref{tab:tab1} provides summary information about the street networks used in this work, while Fig. \ref{fig:spat} shows a visually appealing representation of the networks. For comparison purposes, we also examine a regular 2D lattice grid with 1000 nodes, and each node has degree 4, except the nodes at the boundary of the network.

\section{Results}
\subsection{Model $M_c$ : Constant parameters}
Firstly, we consider all model parameters to be a constant {\sl independent} of nodal properties, {\it i.e.}, \( n^*_i = n^* \), \( \mu_i = \mu \), \( p_i = p \), and \( r_i = r \) for all the nodes on the network. In subsequent sections, we shall denote as $M_c$, the model with parameters independent of nodal properties. The parameter values are fixed at \( \mu = 0.2 \), \( n^* = 10 \), and \( r = 1 \), and the behavior of the congestion parameter $\rho$ as a function of the birth probability \( p \) is investigated as a function of congestion control parameter \( \bar{\eta} \). 

\begin{figure*}
    \centering
    \includegraphics[width=0.99\textwidth]{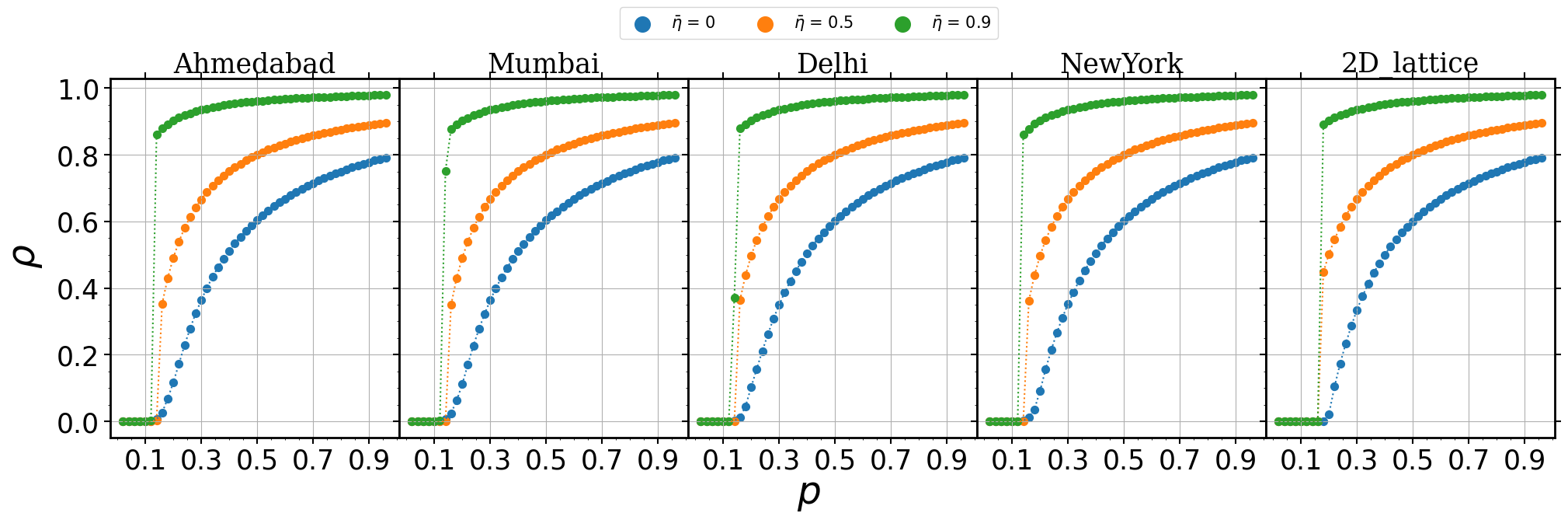}
    \caption{Simulation results for Model-$M_c$ for 4 cities and 2D lattice with parameters fixed at $\mu = 0.2$,  $n^{\ast} = 10$ and $r_i = 1$ for all $i=1,2, \dots N$. Plot shows order parameter $\rho$ as a function of birth probability $p$ for different rejection probability \( \bar{\eta } \).}
    \label{fig:const_PT_1}
\end{figure*}

Figure \ref{fig:const_PT_1} displays the order parameter $\rho$ as a function of birth probability $p$ at three different rejection probability \( \bar{\eta } \). At a coarse level, it is evident that for all the cities considered here a nearly identical behaviour is observed and it is similar to that of a 2D lattice. For the 2D lattice, it is known that the critical birth probability $p_c \simeq \mu$ and is independent of \( \bar{\eta } \) \cite{de2009congestion, de2009minimal}, a feature we observe in Fig. \ref{fig:const_PT_1} as well. For the street networks of cities, despite different structures at microscopic level, we observe that \( p_c < \mu \) at \( \bar{\eta } = 0 \). Further, \( p_c \) decreases slightly as \( \bar{\eta } \) increases. Due to the heterogeneity in the degree of urban street networks, it was expected that \( p_c \) might increase with \( \bar{\eta }\) as observed for scale free network in Ref. \cite{de2009minimal, de2009congestion}. However, our observations reveal an opposite trend: $p_c$ decreases mildly with increase in $\bar{\eta}$.

\subsection{Model $M_d$ : Degree-dependent parameters}
\begin{figure*}
    \centering
     \includegraphics*[width=0.99\textwidth]{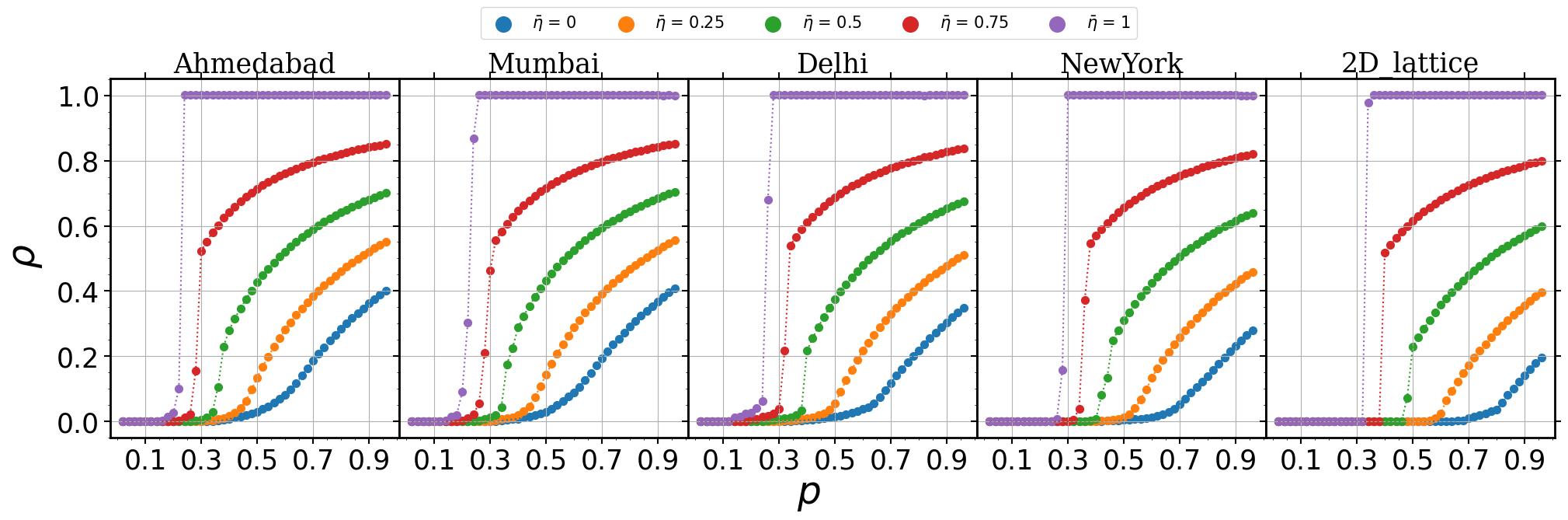}
    \caption{Simulation results for Model-$M_d$ for 4 cities and 2D lattice with parameters fixed at $\mu_i = 0.2$,  $n^{\ast}_i = k_i$ and $r_i = k_i$ for $i=1,2,\dots N$. Plot shows order parameter $\rho$ as a function of birth probability $p$ for different rejection probability \( \bar{\eta } \).}
    \label{fig:Deg_PT}
\end{figure*}

In real traffic, nodal capacity and outflux is usually a function of the degree of the node. Larger the degree of node, larger is its capacity to service the flux of vehicles. Hence, it is natural to modify the basic model by introducing degree dependent outflux and nodal capacity. The parameters are chosen to be $n^*_i = k_i$, $\mu_i = 0.2$, $p_i=p$, and $r_i=k_i$ for all nodes, where $k_i$ represents the degree of $i$-th node. We shall refer to this as Model $M_d$. For these choice of parameters, Fig. \ref{fig:Deg_PT} shows the simulated results for $\rho$ as a function of $p$. At a gross level, this exhibits qualitatively similar features as observed in the case of model-$M_c$, depicted in Fig. \ref{fig:const_PT_1}. Due to the degree dependence of $n^*$ and $r_i$, both nodal capacity and outflux have increased. As a result, network allows more free-flow regime leading to delay of congestion onset in parameter space. Thus $p_c$ is larger than for model-$M_c$.

Additionally, $p_c$ becomes inversely proportional to \( \bar{\eta } \). Consequently, escalating congestion levels deteriorate the protocol's performance on planar networks. Conversely, on scale-free networks, increasing $\eta$ enhances the protocol's efficacy, at least for constant parameter values, as $p_c$ rises with \( \bar{\eta } \) \cite{de2009congestion,de2009minimal}. We performed simulations of model-$M_d$ on scale-free networks. Surprisingly, we find that congestion is entirely eradicated, thus $\rho(p) = 0$ for any \( \bar{\eta } \)  \cite{de2009minimal, de2009congestion,agarwal2023ee}. Considering that real city networks exhibit some heterogeneity akin to scale-free networks, it was anticipated that the critical birth probability $p_c(\bar{\eta } )$ for cities would surpass that of a 2D lattice (a homogeneous network). However, our observations reveal an opposite trend. 

In Fig. \ref{fig:Deg_PT}, for $ \bar{\eta } \gtrsim 0.5$, a gradual transition from free-flow to partially congested state is observed, followed by a sudden, discontinuous transition to congested regime. This phenomenon is prominently observed in the urban traffic dynamics of Ahmedabad, Mumbai, and Delhi, while it exhibits lesser significance in New York and is negligible in a 2D lattice.

\subsection{Model $M_{dr}$ : scaled outflux with rejection probabiity}
Now, we consider a realistic feature that whenever a node or junction overshoots its capacity, the incoming flux is completely rejected. That is, we apply the rule
\begin{equation}
\bar{\eta } = 1, ~~~{\rm if} ~~~n_i > n_i^*.    
\end{equation}
We assume that vehicles are generated on each node at each time step, {\it i.e.}, $p=1$. Other parameters are scaled by $a_0>0$ and $b_0>0$, and are taken to be $\mu=0.2$, $r_i=b_0k_i$ and $n^{\ast} = a_0k_i$. Note that the outflux cannot be greater than the nodal buffer, implying that $a_0 \geq b_0$. In this setting, it would be useful to map the parameter values $a_0$ and $b_0$ for which extreme events can be observed on planar networks. For different combinations of $a_0$ and $b_0$, we find mainly four distinct behaviour of $A(t)$ with time $t$: ({\it i}) linear variation, ({\it ii}) piece-wise linear, ({\it iii}) congested and free-flow coexist and ({\it iv}) free-flow regime. These four cases can be grouped into three regimes, (a) congested (includes (i)), (b) weakly congested (includes cases (ii) and (iii)), and (c) free flow (includes case (iv)). This classification is also supported by the order parameter values calculated for each $(a_0,b_0)$ pair. The congested state refers to $\rho = 1$, while weakly congested state has $0 < \rho < 1$, and for free-flow $\rho = 0$.

Figure \ref{fig:Delhi_ab_TimeS} illustrates the behaviour on the Delhi city network system. 
Recall the condition that $a_0 \ge b_0$. Irrespective of $a_0$, if $b_0$ is sufficiently small, then $A(t) \propto t$ and this corresponds to congested regime having $\rho=0$, shown in Fig. \ref{fig:Delhi_ab_TimeS}(a). For a fixed value of $a_0$, as $b_0$ increases the system either experiences weak congestion or coexists in a state of free-flow and low congestion, as indicated by $0 < \rho < 1$ and can be seen in Fig. \ref{fig:Delhi_ab_TimeS}(b). For large values of $a_0$ and $b_0$, a free-flow regime exists, $\rho=0$), as seen in \ref{fig:Delhi_ab_TimeS}(c). Qualitatively similar behaviour is observed for other cities that had organic growth such as Mumbai and Ahmedabad.

In planned urban regions such as Manhattan in New York city, the weak congestion regime is absent. This is observed in Fig. \ref{fig:NY_ab_TimeS}.  Irrespective of $a_0$, if $b_0$ is sufficiently small, $A(t) \propto t$ and a congestion phase emerges as seen in Fig. \ref{fig:NY_ab_TimeS}(a).  If $b_0$ is moderately high or high, we observe free-flow state as observed in Fig. \ref{fig:NY_ab_TimeS}(b,c). In particular, weakly congested state (encountered for Delhi in Fig. \ref{fig:Delhi_ab_TimeS}(b)) is not observed. This feature -- absence of weakly congested state -- is observed in a 2D grid network as well. This is to be expected since New York (Manhattan) network has a grid like structure similar to a 2D network. 

\begin{figure}[t]
    \centering
   \includegraphics[width=0.5\textwidth]{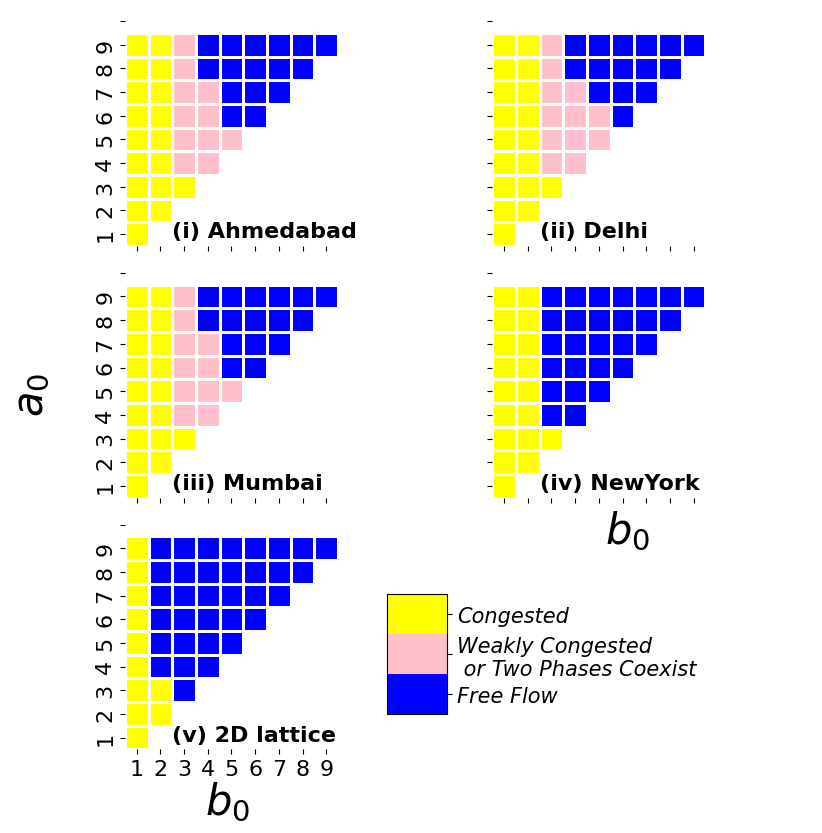}
    \caption{Heat map showing the phases of congestion for different values of $a_0$ and $b_0$, with $p=1$, $\mu = 0.2$,  $n^{\ast} = a_0k_i$, $r_i = b_0k_i$, and $\bar{\eta}$ = 1, for four cities and 2D lattice (grid). Notice the similarity of the heat map for New York (Manhattan) and 2D lattice.}
    \label{fig:Heat_map}
\end{figure}

To obtain a broader picture, we compute the order parameter $\rho$ (in Eq. \ref{eq:orderparam}) for the 4 cities and 2D lattice by varying  $a_0$ and $b_0$ over a larger range. The results are displayed in Fig. \ref{fig:Heat_map} as a heat map. The pattern is clearly visible; in all the cities and 2D lattice shown in Fig. \ref{fig:Heat_map}, a congested phase exists for any $a_0$ and small values of $b_0$. It appears that weakly congested regime emerges for moderate values of $a_0$ and $b_0$ only in cities that have had unplanned but organic growth, as seen in Fig. \ref{fig:Heat_map}(i-iii). The exception is New York city (Manhattan) and 2D lattice (Fig. \ref{fig:Heat_map}(iv-v)). In all the cases, for large values of $a_0$ and $b_0$, free flow regime emerges. In general, time taken for queues to be cleared is larger for congested regime compared to the free-flow regime.

\begin{figure}
    \centering
    \begin{tikzpicture}
        \node[anchor=south west,inner sep=0] (image) at (0,0) {\includegraphics[width=0.32\textwidth]{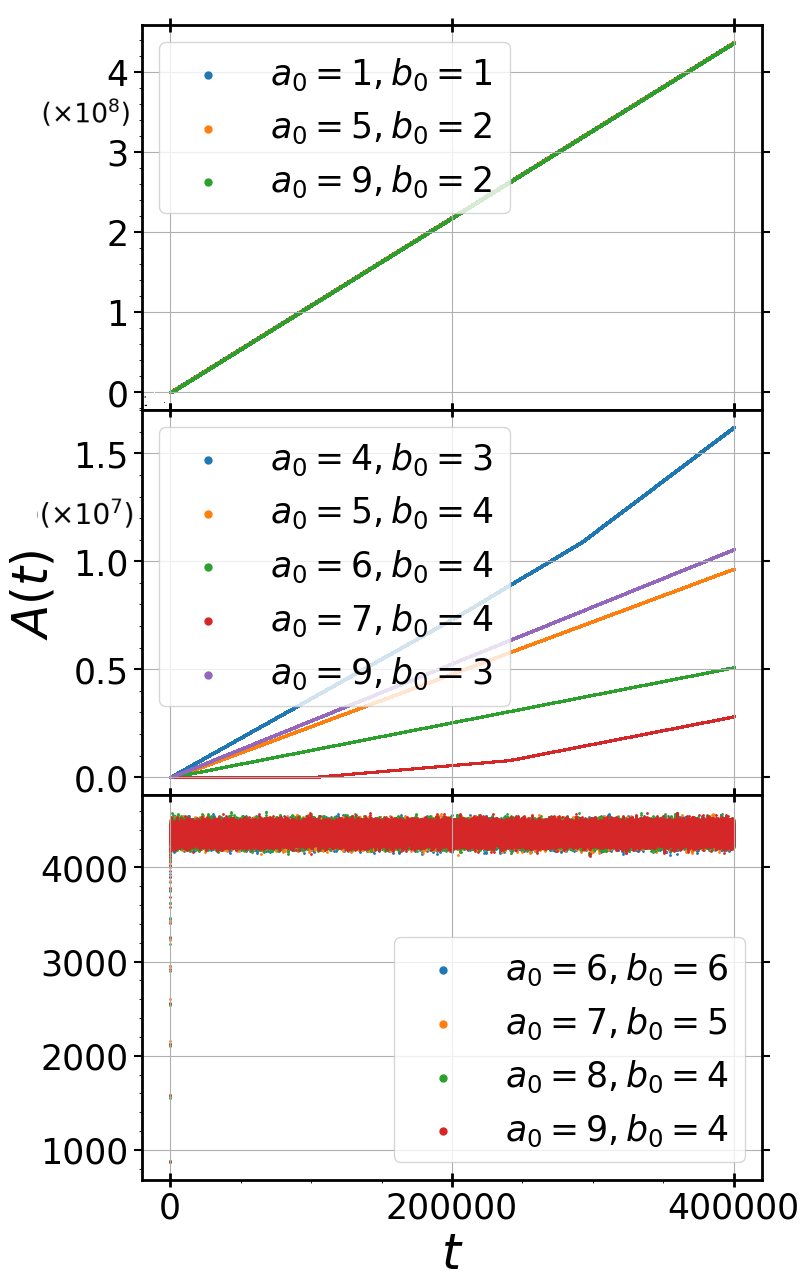}};
        \begin{scope}[x={(image.south east)},y={(image.north west)}]
            \node[anchor=center] at (0.28,0.80) {\textbf{(a)}};
            \node[anchor=center] at (0.28,0.45) {\textbf{(b)}};
            \node[anchor=center] at (0.28,0.2) {\textbf{(c)}};
        \end{scope}
    \end{tikzpicture}
    \caption{Dynamics on Delhi street network for various values of $a_0$ and $b_0$. (a) Congested phase : For small $b_0$, and regardless of $a_0$. $A(t) \propto t$ indicating congestion with $\rho = 1$. (b) Weak congestion : For moderate values of $a_0$ and $b_0$, weak congestion ($0< \rho < 1$) or coexistence of free-flow and low congestion encountered. (c) Free flow regime : For large $a_0$ and $b_0$, free-flow exists with $\rho = 0$.}
    \label{fig:Delhi_ab_TimeS}
\end{figure}
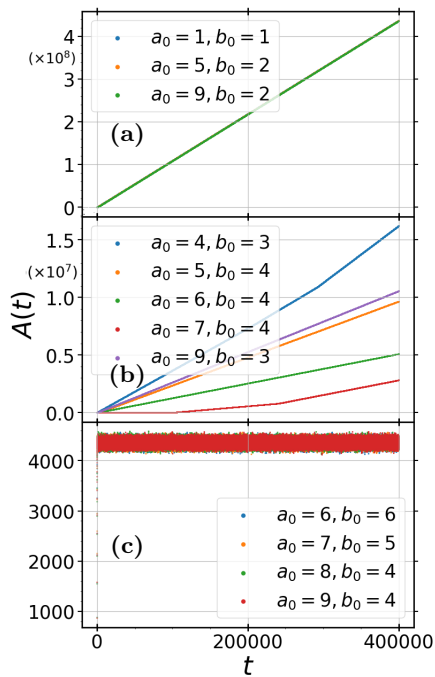
\begin{figure}
    \centering
    \begin{tikzpicture}
        \node[anchor=south west,inner sep=0] (image) at (0,0) {\includegraphics[width=0.32\textwidth]{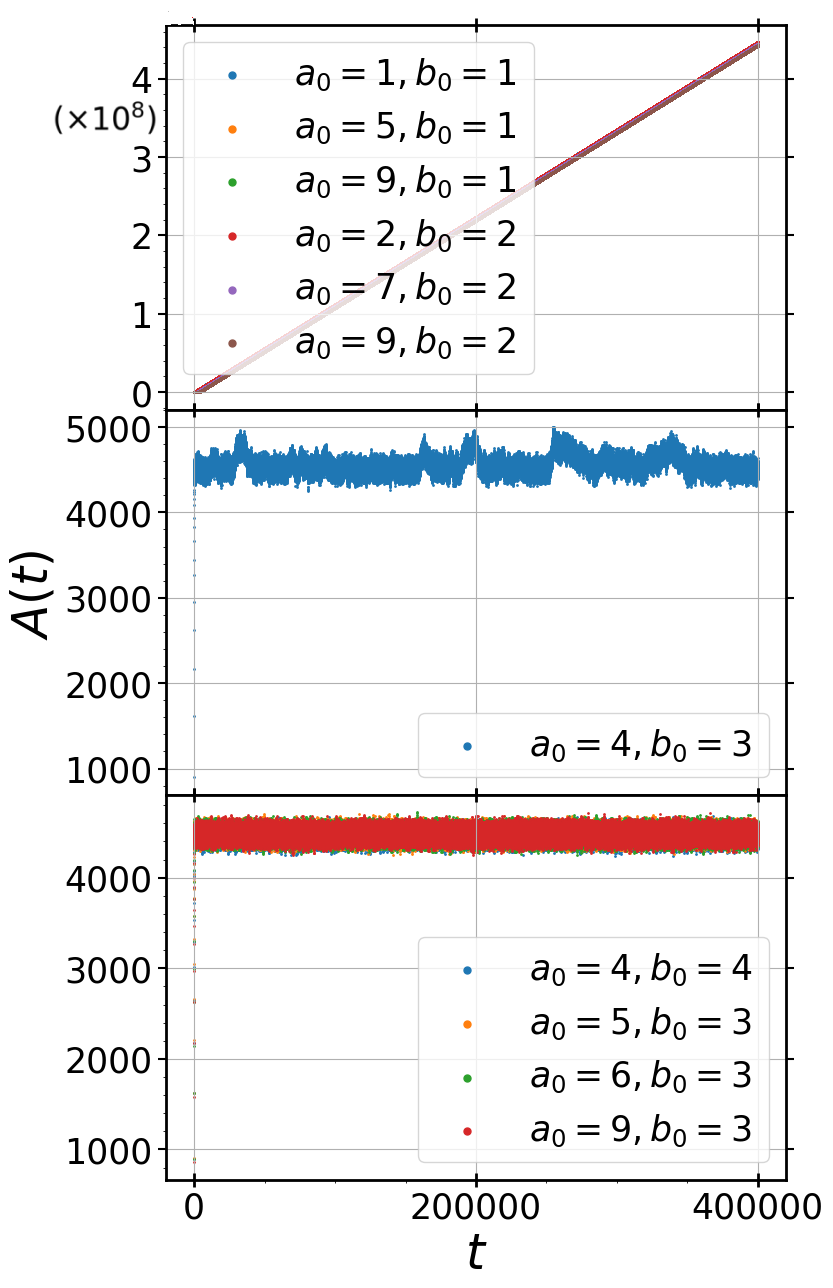}};
        \begin{scope}[x={(image.south east)},y={(image.north west)}]
            \node[anchor=center] at (0.28,0.80) {\textbf{(a)}};
            \node[anchor=center] at (0.28,0.45) {\textbf{(b)}};
            \node[anchor=center] at (0.28,0.2) {\textbf{(c)}};
        \end{scope}
    \end{tikzpicture}
    \caption{Dynamics on New York street network. (a) Congested phase with $b_0 \leq 2$, regardless of $a_0$ (b) Free flow phase with \(a_0 = 4\) and \(b_0 = 3\), (c) Free flow phase for large values of $a_0$ and $b_0$. In (b), \(\Delta t\) is larger compared to the free-flow regime shown in (c), which is expected as states with \(a_0 < 4\) and \(b_0 < 3\) correspond to congested regions (see Fig. 8(iv)).}
    \label{fig:NY_ab_TimeS}
\end{figure}
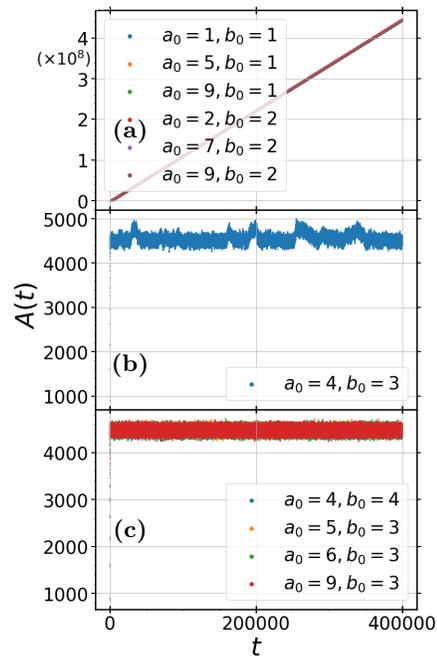

\subsection{Extreme events in free-flow regimes}
For the present problem, an ``event'' at $i$-th node would be the number vehicles $n_i$ passing through the node. 
An extreme event is the one for which the flux strongly deviates from the mean flux through that node. Hence, an extreme event at time $t$ must satisfy the condition $n_i(t) > q$, where $q$ is the node-dependent threshold given by \cite{kishore2011extreme, kishore2012extreme},
\begin{equation}
    q_i~= ~\langle n_i \rangle +~m\sigma_i,
\label{eq:threshold}
\end{equation}
where $\langle n_i \rangle$ is the average flux and $\sigma_i$ is standard deviation of the flux, and $m \ge 0$ is a real number. The notion of extreme events requires that the flux through nodes be stationary, i.e, the distribution $n$ must be time-independent. Hence, the extreme event value can be computed only for the free-flow state. The probability of occurrence of an extreme event on node $i$ is given by, 
\begin{equation}
    P_{EE}^i = \frac{1}{T}\sum_{t=0}^{T} \theta(n_i(t) -  q_i),
\end{equation}
where $T$ is the number of time steps and $\theta(x)$ is a Heaviside function.

For the set of parameters $p=1$, $\mu=0.2$, $\eta = 1$, $r=b_0k_i$ and $n_i^\ast = a_0k_i$, free-flow state  emerges at $a_0=6$ and $b_0=6$ for Ahmedabad, Delhi and Mumbai. For grid-like structure of New York (Manhattan) and 2D lattice, a free-flow state emerges at $a_0=4$ and $b_0=4$.

\begin{figure}[t]
    \centering
   \includegraphics[width=0.4\textwidth]{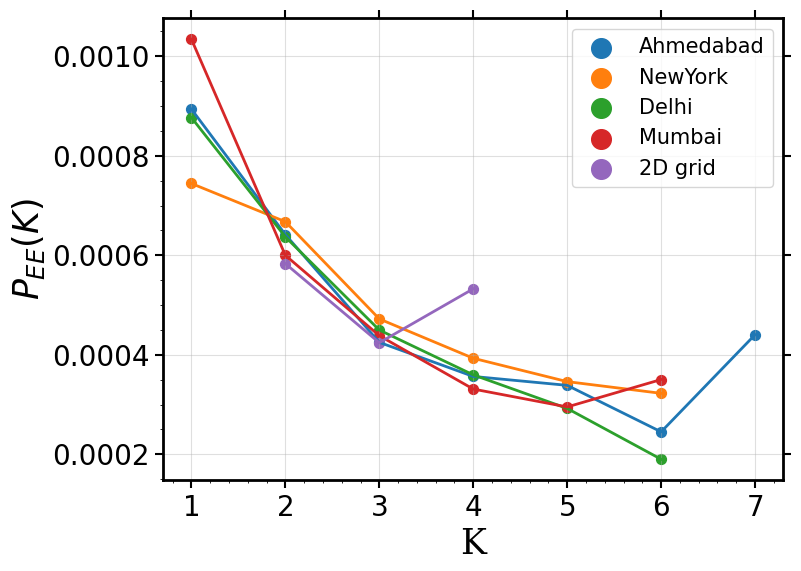}
    \caption{Probability for the occurrence of extreme events $P_{EE}$ as a function of degree $k$ on urban street networks and 2D square grid.}
    \label{fig:EE_deg}
\end{figure}

The probability for the occurrence of an extreme event on node $i$ is given by, 
\begin{equation}
    P_{EE}^i = \frac{1}{T}\sum_{t=0}^{T} \theta(n_i(t) -  q_i)
\end{equation}
where $n_i(t)$ represents the number of walkers on node $i$, $T$ is the number of time steps in the time series, $q_i$ threshold for node $i$ and $\theta(x)$ is a Heaviside function.
We compute extreme event probability $P_{EE}^i$ as function of degree $k$ of nodes and is displayed in Fig. \ref{fig:EE_deg}. In this figure, $P_{EE}^i$ is averaged over all the nodes with same degree. It is evident that, for all the planar networks in Fig. \ref{fig:EE_deg},  extreme event probability is higher for small degree nodes than in the hubs. This is surprising, considering that hubs attract more vehicles, and despite this, the extreme events occurrence probability on them is smaller than that for small degree nodes. This is reminiscent of a similar result for random walks on scale-free (non-planar) networks \cite{kishore2011extreme}. Thus, the central feature -- smaller extreme event probability on hubs -- is valid even if the dynamics is not governed by simple random walks and the networks are not planar. Furthermore, we note that the extreme events have the same order of magnitude for any $(a_0, b_0)$ pair, provided the planar system is in a free-flow regime.

\section{Summary}
In this work, we set out to study the dynamics of a realistic traffic model on urban street (planar) networks of four cities and, for comparison, on a two-dimensional regular grid. The central aim is to understand the main features of the occurrence of congestion and extreme events on street networks. We adopted the random walk model dynamics with several additional realistic features such as birth and death rates for vehicles and FIFO queues to study congestion phenomenon, phase transition and extreme events. 

Broadly, the traffic dynamics on urban street networks qualitatively display identical behaviour to that of 2D regular grid network as far as congestion and extreme events are concerned. Despite each street network of cities being vastly different, the results reveal that the nature of phase transition from free-flow to congested regime as a function of birth rate of vehicles is similar for all the street networks and square grid. We demonstrate this using three variants of the random walk based dynamical model, each variant differing in terms of how the influx to a node and outflux from a node are parameterised. \\[1mm]

Further, we provide a broad overview of the parametric regimes in which free flow and congested phases emerge. We also uncover existence of a semi-congested regime (congestion and free-flow phases co-exist) in cities with organic and unplanned growth. This phase is absent in planned regions of cities such as New York (Manhattan) and in the 2D regular grid. Finally, we show that extreme events have a higher probability of occurrence on small degree nodes of the street networks compared to that on hubs.  In future, these initial results on four urban networks must be verified on a larger sample of urban networks, and the validity of the main results related to congestion and extreme events must be tested against real traffic data from street networks and through indirect proxies such as mobile phone traces \cite{CcoLimGon2016}. This would help understand the dynamics of congestion and extreme events in a realistic scenario, and ultimately provide inputs for designing measures towards controlling them.

\section{Acknowledgment}
Ajay Agarwal extends sincere gratitude to IISER, Pune for their generous funding support. The numerical simulations presented in this study were performed on the PARAM Brahma supercomputer at IISER, Pune. We thank Aanjaneya Kumar and Ritam Pal for their valuable contributions to the discussions.

\bibliography{Bibliography}
\end{document}